\documentclass{icrc}

\usepackage{times}
\usepackage{graphicx} 

\begin{document}

\title{A new airborne detector for atmospheric muons}
\author[1]{C.P. Achenbach}
\author[1]{J.H. Cobb}
\affil[1]{ Nuclear and Particle Physics Sub-Department, 
	University of Oxford, 1 Keble Road, Oxford, OX1 3RH, United 
	Kingdom.}

\correspondence{Dr Carsten Patrick Achenbach 
(p.achenbach@physics.ox.ac.uk)}

\firstpage{1}
\pubyear{2001}

\maketitle

\begin{abstract}
  The University of Oxford has started the design and development of
  the new experiment ADLER (\underline{A}irborne \underline{D}etector
  for \underline{L}ow \underline{E}nergy \underline{R}ays). This
  apparatus will measure the cosmic-ray muon flux at an altitude of 10
  -- 13\,km. The detector should be flown by aircrafts on transatlantic
  routes crossing the magnetic equator to investigate the flux at
  different geomagnetic latitudes. The goal of the experiment
  is to obtain better constraints on the low energy atmospheric
  neutrino flux and the results will be of importance to the
  atmospheric neutrino anomaly.
\end{abstract}

\section{Introduction}
 
One of the most exciting results to emerge in particle physics in the
last decade is the apparent deficit of atmospheric muon neutrinos
reported by underground experiments (Fukuda et al., 1998; Ambrosio et
al., 1998; Allison et al., 1998). The latest results from the Japanese
Super-Kamiokande experiment provide strong evidence for a flavour
oscillation of muon neutrinos into tau neutrinos, but the precision is
limited by uncertainties in the predicted flux of atmospheric
neutrinos with energies less than 1\,GeV. The neutrinos are end
product of a cascade of particles in the atmosphere: Incident cosmic
rays (CR) produce light mesons in nuclear collisions with air
molecules. Charged pions decay to muons and neutrinos; below 2.5\,GeV
the majority of muons will themselves decay to electrons or positrons
and neutrinos. Charged kaons show a similar decay chain:
\begin{eqnarray*}
  A_{\mbox{CR}} + A_{\mbox{Air}} & \rightarrow & \pi^\pm, \pi^0,
  K^\pm, \mbox{other hadrons}\\ \pi^\pm, K^\pm & \rightarrow &
  \mu^\pm\ + \buildrel \hspace{-1mm}(-) \over {\nu_\mu}\\ \mu^\pm\ &
  \rightarrow\ & e^\pm\ + \buildrel \hspace{-1mm}(-) \over {\nu_e} +
  \buildrel \hspace{-1mm}(-) \over {\nu_\mu}
\end{eqnarray*}

The uncertainties of present Monte Carlo calculations performed by the
Bartol group (Barr et al., 1989; Agrawal et al., 1996) and by Honda,
Kajita, Kasahara and Midorikawa (Honda et al., 1990) can be reduced by
measuring the flux of low-energy muons. Those muons are strongly
correlated to neutrinos but much easier to detect. While muon
measurements at ground level are widely reported in the literature,
there have been very few attempts to measure the muon flux as a
function of altitude and latitude. An early experiment performed by
Conversi (1950) gave the first information about muons at aircraft
altitudes. The deployment of balloon-borne detectors since 1989
allowed more accurate measurements during their ascent in the
atmosphere, but the main difficulty in such experiments remains the
fixed location and the altitude of the detectors is continuously
varying.  A compilation of reported muon flux results on balloon
measurements can be found in Ambriola et al., 2000.

This paper reports on the design of a compact and mobile apparatus to
be flown by an aircraft. The method employed for distinguishing muons
from other ionising particles is the observation of delayed
coincidences in an active absorber together with
anti-\-co{\-in\-ci\-dences in surrounding veto-counters. The detector
concept is similar to Conversi's, although scintillators are used in
place of Geiger-M\"uller tubes.

\section{Instrument description}

\begin{figure*}[t]
  \figbox*{}{}{\includegraphics[width=11.0cm]{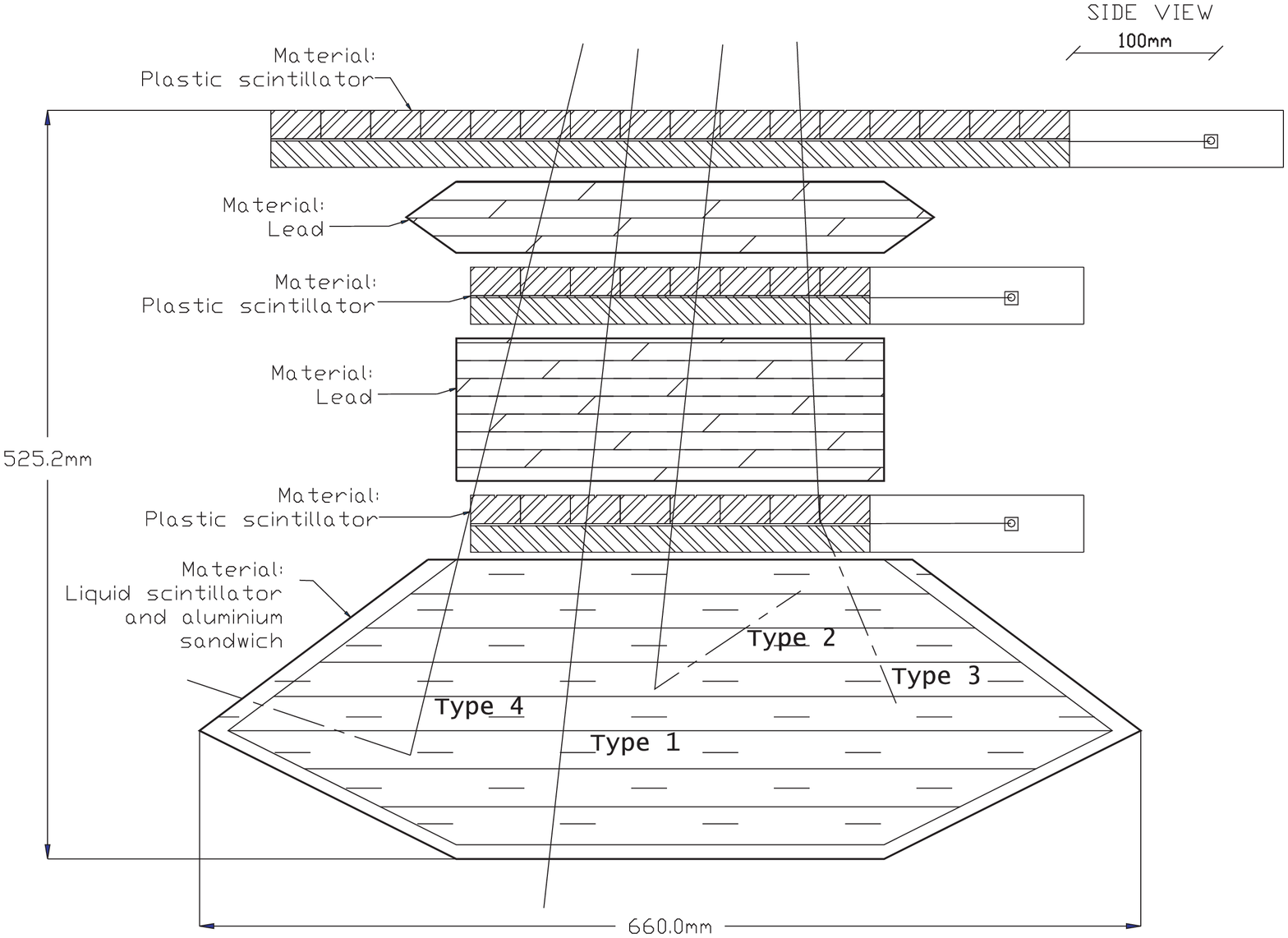}}
  \caption{Schematic drawing of the ADLER detector with a trapezoidal
	active absorber. Different types of particle tracks are 
	distinguished by the timing of the detector response. Tracks 
	of type 1 originate from high energy muons, type 2 tracks
	originate from low energy muons which have stopped in the 
  	active absorber, type 3 tracks originate from muons which 
	have stopped in the third trigger hodoscope and the decay 
	electrons of type 4 tracks produce additional signals in one 
	of the veto-counters and leave the active volume.}
\end{figure*}

The challenge is to construct a detector which is sufficiently rugged,
light and low powered to be carried on an aircraft. Individual
detector parts will be mounted in an steel frame and will be
supported by strong wings in the four corners. The height of the frame
will be 60\,cm and its cross-section 85 $\times$ 85\,cm$^2$. The total
weight including the inactive absorbers is about 350\,kg.

Fig.~1 presents a schematic cross-sectional view of the main
components. The detector will consist of three similar scintillator
hodoscopes. Each hodoscope is composed of two planes of plastic
scintillator bars. For the first one (S1) there are 16 strips of
length 56\,cm in each layer, while 8 strips of length 28\,cm make up
the second (S2) and third (S3). All strips are 3.5\,cm wide by 2\,cm
thick. The maximum detectable zenith angle is defined by the
hodoscopes S1 and S3 and amounts to 53.6$^\circ$. The opening angle
and the active areas define the geometrical acceptance but muons with
larger angles can scatter into the acceptance region.

Two blocks of lead (30\,cm length and width) are stacked between the
scintillators. The bottom part of the detector consists of an active
absorber. The shape of the active absorber is a double trapezoid with
a cross-section of 30 $\times$ 30\,cm$^2$ at top and bottom and a
cross-section of 62 $\times$ 62\,cm$^2$ in the middle. Also different
shapes like double cones or rectangular pyramids have been
investigated. The use of liquid scintillator as the active detector
element, in conjunction with wavelength-shifting fibres for light
collection and transmission, is well suited for such an experiment.
The anticipated commercial liquid scintillator\footnote{Bicron,
Newbury, OH, USA.} \addtocounter{footnote}{-1} BC-517 L contains about
30\% pseudocumene, has a light yield of 39\% anthracene and a flash
point of 102$^\circ$ Celsius. A volume of 60 litres will be contained
in a single aluminium container segmented by optical
separators. Furthermore it is chemically compatible with the blue to
green wavelength-shifting fibre\footnotemark\ BC-91 A (Young et al.,
1993). Fibres of 1.2\,mm diameter will be used in the shape of
Archimedes' spirals inside the tank with an equal distance of 4\,cm
between adjacent turns.

The signals of the scintillators will be read out by the optical
fibres to five small (3 $\times$ 3\,cm) Hamamatsu multi-anode M16
(R5900) photomultiplier tubes operated at voltages of about
1000\,V. The total number of analogue channels is 96. The signal
processing will be done by simple, low-powered electronics and a data
acquisition system based on laptop-PC. The trigger mode will be a
coincidence of signals from the three scintillator hodoscopes which
can be interpreted as a charged track, together with an
anti-coincidence of the veto-counters. The timing of subsequent hits
in the active elements is recorded in a 16 $\times$ 16 $\times$ 16
FIFO using a 40\,MHz clock to keep the rate of false coincidences
low. The time difference between trigger signal and delayed signal in
measures the decay time of the muon. The background to the decay curve
can be restricted by requiring an anti-coincidence of the
veto-counters with the delayed signal if the reconstructed track is of
type~2, see Fig.~1. The power consumption is about 30\,W for the
electronics and 20 -- 30\,W for the laptop. At the present stage of
the experiment the design of the electronics is not yet finished. The
position and altitude of the aircraft at any time could be obtained
from mobile Global Positioning System (GPS) equipment or the aircraft
log.

\section{Muon charge ratio considerations}

\begin{figure}[t]
  \includegraphics[width=8.3cm]{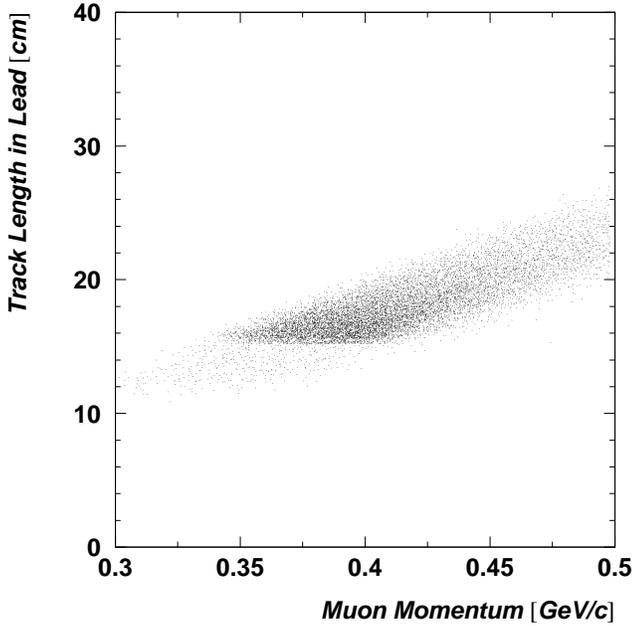} 
  \caption{Histogram of muon track lengths in lead versus muon 
	momentum. Vertical muons travel through 15\,cm of lead, 
	equivalent to 26 radiation lengths.}
\end{figure}

\begin{figure}[t]
  \includegraphics[width=8.3cm]{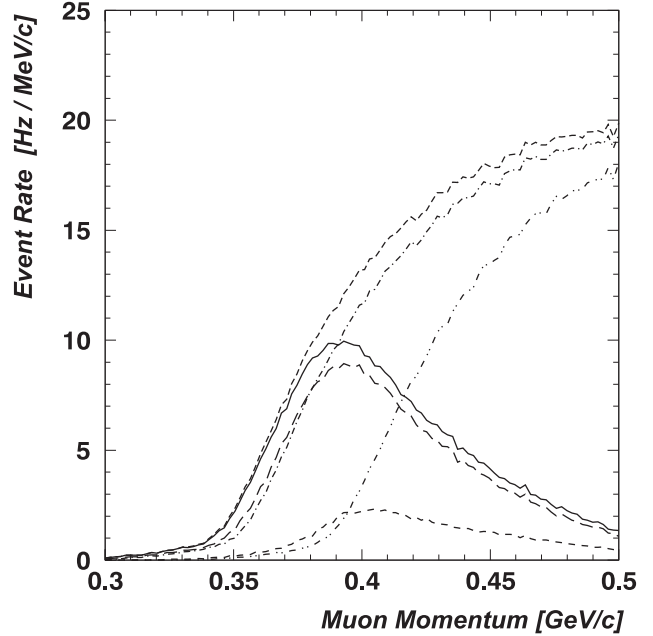} 
  \caption{Acceptance curves of the ADLER detector. The event 
  rates are encoded as follows: dashed = trigger ($S_1S_2S_3$); 
  dot-dashed = muon signals ($S_1S_2S_3A$); dot-dot-dashed = 
  signals in veto-counters ($S_1S_2S_3V$); solid = stopped muons; long 
  dashes = muon decay signals in absorber ($S_1 S_2 S_3 A 
  \overline{V} D_A$); short dashes = muon decay signals in 
  veto-counters ($S_1 S_2 S_3 \overline{V}\! A D_V$).}
\end{figure}

\begin{figure}[t]
  \includegraphics[width=8.3cm]{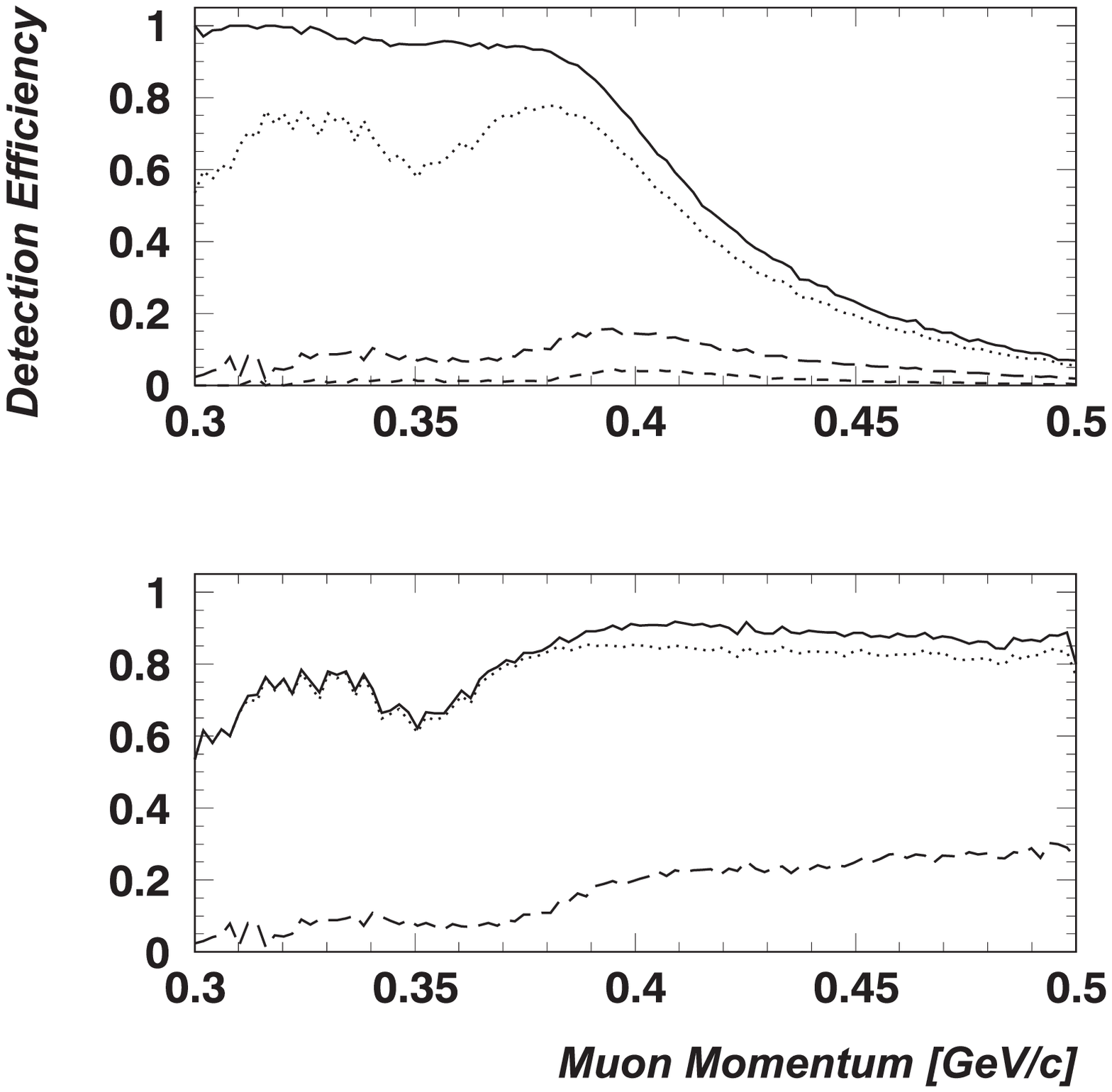} 
  \caption{Efficiency curves of the ADLER detector. The 
  efficiencies are encoded as follows in the upper plot: 
  solid = stopped muons per trigger; dots = muon decay 
  signals in absorber per trigger ($S_1 S_2 S_3 A \overline{V}
  D_A$); long dashes = muon decay signals in veto-counters 
  per trigger ($S_1 S_2 S_3 \overline{V} A D_V$); short 
  dashes = muon decay signals in veto-counters only per 
  trigger ($S_1 S_2 S_3 \overline{V} A \overline{D}_A D_V$); 
  in the lower plot: solid = muon decay signals per stopped muon 
  ($S_1 S_2 S_3 A \overline{V} D$); dots = muon decay signals in 
  absorber per stopped muon ($S_1 S_2 S_3 A \overline{V} 
  D_A$); long dashes = muon decay signals in veto-counters 
  per stopped muon ($S_1 S_2 S_3 A \overline{V} D_V$).}
\end{figure}

The flux ratio $r = N_{\mu^+}/N_{\mu^-}$ of positive to negative muons
provides important information on the interactions of the primary
cosmic rays with nuclei in the atmosphere. The positive excess is
approximately 1.25 in the low energy range. Most of the previous
measurements have been carried out using magnetic deflection to
determine the muon charge with simultaneous determination of the
energy.

The principle of the ADLER measurement is based on the capture of
negative muons according to the reaction $\mu^- + p \to n +
\nu_\mu$. Practically the entire energy liberated in this process is
carried off by the neutrino and no signal will be produced in the
detector. As a result the mean negative muon lifetime becomes $1/\tau
= \Lambda_c + 1/\tau_0$ where $\Lambda_c$ is the capture probability
and $\tau_0$ the decay lifetime for a free muon. At $Z = 11 - 12$ the
capture probability is almost equal to the decay probability, for iron
($Z$ = 26) it is 90\% of the total disappearance probability. For that
reason the active absorber is designed flexible, making measurements
with different absorbers (aluminium, steel, scintillator) possible. It
could consist of horizontal partitions with half of the mass made of
thin (1\,mm) plates and the other half made of liquid scintillator
cells. In the case of steel plates the total fraction of decaying
muons will be 55\%. Thus, the determination of the muon charge ratio
will be achievable from a comparison of total count
rates. Alternatively it can be extracted from the measured decay
curve, which is a superposition of the different lifetimes of negative
muons in liquid scintillator and the absorber material and of positive
muons.

\section{Muon flux considerations}

The mixed radiation field induced by cosmic rays consists of a broad
spectrum of different charged particles with varying momentum. The
expected composition at 14\,km height (atmospheric depth $\approx$ 200
g$/$cm$^2$) is 60\% soft electrons, 30\% protons, 10\% penetrating
muons, 1\% penetrating electrons and 0.1\% pions. The momentum
acceptance and detection efficiencies have been studied carefully
using the GEANT~3.21 Monte Carlo code. The geometrical set-up used was
based on the construction drawings. The total amount of material above
the active absorber amounts to 26 radiation lengths corresponding to
about one nuclear interaction length. This will provide a considerable
shielding of the lower scintillators from soft electrons and
gamma-rays. Fig.~4 shows a histogram of simulated muon track lengths
in the lead blocks of ADLER versus muon momentum.

The most interesting muons are those created with energies around
$E_\mu \approx 2 E_\nu$ and the most important range of neutrino
energy is $E_\nu \approx 0.1 - 10$\,GeV. The mean energy loss of
vertical minimal ionising muons will be about 40\,MeV in the
scintillators and 187\,MeV in the lead blocks. In Fig.~2 the
acceptance curves are plotted versus muon momentum. Different detector
signals have been calculated: Muons with momentum above 300\,MeV$/$c
will trigger the data acquisition (dashed line). The veto-counters
will give a signal (dot-dot-dashed line) for penetrating muons. The
rate of muons with identified decays is given by the coincidence
requirements and is shown by the long dashed curve. This rate has to
be compared with the total rate of stopped muons in the simulation
(solid line). The FWHM of the acceptance curve for identified muons is
61\,MeV$/$c corresponding to an effective momentum width of
28\,MeV$/$c for the nominal geometrical factor of
2088\,cm$^2$\,sr. This factor has been calculated by means of two
independent codes and allows to estimate the expected flux of muons of
one charge for a given geomagnetic latitude. Assuming a flux of
0.01\,$\mu/$(cm$^2$\, sr\,s\,GeV$/$c) at a geomagnetic cut-off of
4.5\,GV gives a count rate of 0.6\,Hz or 2,200 counts per hour. A
measurement with 3\% statistical error can be performed in about
1/2\,h. During this time a south-going flight covers a geomagnetic
latitude difference of 4$^\circ$. Those numbers lead to the expected
count rates listed in Table~1.

\begin{table}[ht]
  \caption{Expected count rates in the detector components. The first 
	line of each component presents the correlated background rate, 
	every other line presents the uncorrelated background rate.}
  \begin{center}
    \footnotesize
    \begin{tabular}{lrrr}
    \hline\\[-3mm]
    Counter & \multicolumn{1}{c}{Acceptance} & \multicolumn{1}{c}
	{Flux $\times$ 10$^{-2}$} & \multicolumn{1}{c}{Rate} \\
	& \multicolumn{1}{c}{[cm$^2$\,sr]} & \multicolumn{1}{c}
	{[(cm$^{2}$ s sr)$^{-1}$]} & \multicolumn{1}{c}{[Hz]} \\
    \hline \hline
    Scintillator S1 	&  2000 & 30 \hspace{5mm} & 2 $\times$  600 \\
 			&  8000 & 30 \hspace{5mm} & 2 $\times$ 2400 \\
    Scintillator S2	&  2000 & 10 \hspace{5mm} & 2 $\times$  200 \\
    	   		&   400 & 30 \hspace{5mm} & 2 $\times$  120 \\
    Scintillator S3 	&  2000 &  8 \hspace{5mm} & 2 $\times$  160 \\
    		   	&   400 & 30 \hspace{5mm} & 2 $\times$  120 \\
    Active absorber 	&  2000 &  8 \hspace{5mm} & 8 $\times$  160 \\
     (one cell)   	& 10000 & 30 \hspace{5mm} & 8 $\times$ 3000 \\
    Veto-counters   	&  2000 &  8 \hspace{5mm} & 160  \\
    		   	& 10000 & 30 \hspace{5mm} & 3000 \\
    \hline
    ($S_1 S_2 S_3$)     	& & & \bf  120  \\
    ($S_1$ or $S_2$ or $S_3$)	& & & \bf 7200  \\
    \hline
    \end{tabular}
  \end{center}
\end{table}

Fig.~3 shows the corresponding efficiency curves, where the upper plot
refers to count rates normalized to the trigger rate and the lower
plot refers to count rates normalized to the rate of stopped
muons. The efficiency of detecting a stopped muon is about 90\%. In
addition, 20\% of the decay electrons are detected in the
veto-counters providing a redundancy in the data analysis which can be
used to reduce backgrounds.

\section{Aircraft considerations}

\balance The effect of the geomagnetic field is recognized as very
important in atmospheric neutrino flux calculations. At present, the
confidence in the IGRF (International Geomagnetic Reference Field)
models is strong. However, a measurement of the muon flux at different
geomagnetic latitudes $\lambda$ is desirable. The flight paths
investigated involve transatlantic routes. The possible latitude band
extends from $\sim 55^\circ$\,N to $\sim 35^\circ$S.  The majority of
commercial flights have cruising altitudes from 11 to 13\,km.

The whole structure of the detector must tolerate acceleration in each
direction to meet the regulations of the aircraft. In the vertical and
horizontal directions perpendicular to the aircraft axis the limits of
maximum acceleration are at 3$g$. Along the aircraft axis the limit is
at 9$g$. Finite element analysis has been performed during the design
of the detector to meet those demands.

\section{Conclusions}

The ADLER detector is a small and compact apparatus for measuring the
flux of atmospheric muons as a function of the geomagnetic
coordinates. The momentum acceptance was chosen to be relevant to the
atmospheric neutrino anomaly. The muon charge ratio will be determined
by a comparison of the $\mu^-$ lifetimes in different absorbers. The
results will lead to a better understanding of present and future
atmospheric neutrino experiments. Furthermore, the measurements will
provide an important check of the new 3-dimensional atmospheric
neutrino flux calculations being performed at Oxford.


\begin{thebibliography}{99}

\bibitem{Agrawal, 1996} Agrawal, V. et al., Atmospheric neutrino flux
above 1 GeV, Phys. Rev., D53, 1314--1323, 1996.

\bibitem{Allison, 1998} Allison, W.W.M. et al. (Soudan-2 Coll.), The
atmospheric neutrino flavor ratio from a 3.9 fiducial kiloton year
exposure of Soudan-2, Phys. Lett., B449, 137--144, 1998.

\bibitem{Ambriola, 2000} Ambriola, M.L. et al., The WiZard
collaboration cosmic ray muon measurements in the atmosphere,
Nucl. Phys. (Proc. Suppl.), B85, 355-360, 2000.

\bibitem{Ambrosio, 1998} Ambrosio, M. et al. (MACRO Coll.),
Measurement of the atmospheric induced upgoing muon flux using MACRO,
Phys. Lett., B434, 451--457, 1998.

\bibitem{Barr, 1989} Barr, G. et al., Flux of atmospheric neutrinos,
Phys. Rev., D39, 3532--3534, 1989.

\bibitem{Conversi, 1950} Conversi, M., Experiments on cosmic-ray
  mesons and protons at several altitudes and latitudes, Phys. Rev.,
  79(5), 749--767, 1950.

\bibitem{Fukuda, 1998} Fukuda, Y. et al. (Super-Kamiokande Coll.),
Evidence for oscillation of atmospheric neutrinos, Phys. Rev. Lett.,
81, 1562--1567, 1998.

\bibitem{Honda, 1990} Honda, M. et al., Atmospheric neutrino fluxes,
Phys. Lett., B248, 193--198, 1990.

\bibitem{Young, 1993} Young, K.G. et al., Radiation and solvent 
effects on wavelength shifting fibers used with liquid scintillators,
Radiat. Phys. Chem., 41(1--2), 215--219, 1993.

\end{thebibliography}
\end{document}